\documentclass{ws}

\begin{document}

\def\bb{b\bar{b}}
\def\bu{B^+}
\def\b0{B^0} 
\def\bd{B^0_d} 
\def\bq{B^0_q}
\def\bs{B^0_s}
\def\b0b{\overline{B^0}} 
\def\bdb{\overline{B^0_d}} 
\def\bqb{\overline{B^0_q}} 
\def\bsb{\overline{B^0_s}} 
\def\lb{\Lambda_b}
\def\bmix{B^0 \mbox{--} \overline{B^0}}
\def\bdmix{B_d^0 \mbox{--} \overline{B_d^0}}
\def\bsmix{B_s^0 \mbox{--} \overline{B_s^0}}
\def\bsg{b\to s\,g}
\def\dmd{\Delta m_d}
\def\dmq{\Delta m_q}
\def\dms{\Delta m_s}
\def\epem{e^+ e^-}
\def\ips{\mbox{ps}^{-1}}
\def\kmix{K^0 \mbox{--} \overline{K^0}}
\def\kstar{K^{\ast 0}}
\def\kstarbar{\overline{K}^{\ast 0}}
\def\sinsqth{\sin^2\theta_W^{eff}}
\def\ups4s{\Upsilon_{4S}}
\def\vtd{V_{td}}
\def\vts{V_{ts}}
\def\Zbb{Z^0 \rightarrow b\,{\overline b}}
\def\Zcc{Z^0 \rightarrow c\,{\overline c}}
\def\Zff{Z^0 \rightarrow f\,{\overline f}}
\def\Zuds{Z^0 \rightarrow u\,{\overline u},d\,{\overline d},s\,{\overline s}}

\title{A Review of $\bs$ Mixing: Past, Present and Future}

\author{St\'ephane Willocq}

\address{Physics Department, University of Massachusetts, MA 01003,
USA\\E-mail: willocq@physics.umass.edu}


\maketitle

\abstracts{
We review the experimental status of $\bsmix$ mixing.
After a brief historical overview, current studies of
the time dependence of $\bs$ oscillations are described,
with an emphasis on the different experimental techniques used by
the ALEPH, CDF, DELPHI, OPAL, and SLD Collaborations.
To conclude, the outlook for future experiments is presented.
}

\section{Introduction}

In analogy to the $\kmix$ system, the $\bmix$ system consists
of $B^0$ and $\overline{B^0}$ flavor
eigenstates, which are superpositions of heavy and light mass eigenstates
$B_H$ and $B_L$.
Due to the difference in mass and width, the mass eigenstates evolve differently
as a function of time, resulting in time-dependent $\bmix$ flavor oscillations
with a frequency equal the mass difference $\Delta m \equiv m_H - m_L$. As a consequence,
an initially pure $|B^0\rangle$ state may be found to decay as $|B^0\rangle$
or $|\overline{B^0}\rangle$ at a later time $t$ with a probability
density equal to
$P(B^0 \to B^0) = \frac{\Gamma}{2} e^{-\Gamma t} (1+\cos\Delta m\, t)$ or
$P(B^0 \to \overline{B^0}) = \frac{\Gamma}{2} e^{-\Gamma t} (1-\cos\Delta m\, t)$. 
(Here we have taken $\Gamma_L \simeq \Gamma_H \simeq \Gamma$ since
$\Delta \Gamma << \Delta m$ in the Standard Model.)

\begin{figure}[t]
  \vspace*{-4mm}
  \centering
  \epsfxsize=12cm
  \epsfbox{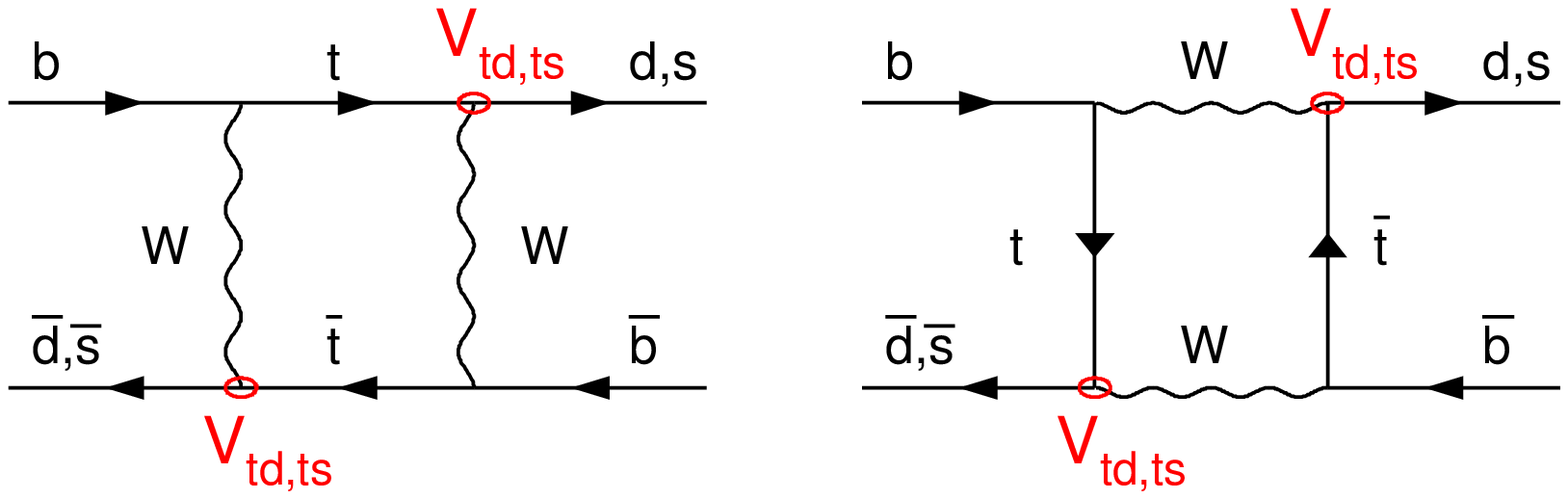}
  \vspace*{-12mm}
  \caption{\it \label{fig_box}
  \baselineskip=12pt
  Box diagrams leading to $\bmix$ mixing. Only the dominant top quark contribution
  is shown.}
  \baselineskip=18pt
\end{figure}

The oscillation frequency $\dmq$ ($q= d$ and $s$ for $\bd$ and $\bs$)
can be computed via the second order
box diagrams that induce $B^0 \leftrightarrow \overline{B^0}$
transitions, see Fig.~\ref{fig_box}.
Calculations yield\cite{Buras}
\begin{equation}
  \dmq = \frac{G_F^2}{6\pi^2} m_{B_q} m_t^2 F(m_t^2 / m_W^2) f_{B_q}^2 B_{B_q}
         \eta_{QCD} \left| V_{tb}^\ast V_{tq} \right|^2,
  \label{eq_dmq}
\end{equation}
where $G_F$ is the Fermi constant, $m_{B_q}$ is the $B^0_q$ hadron mass, $m_t$ is
the top quark mass, $m_W$ is the $W$ boson mass,
$F$ is the Inami-Lim function,\cite{Inami}
and $\eta_{QCD}$ is a perturbative QCD parameter.
The ``bag'' parameter $B_{B_q}$ and the decay constant $f_{B_q}$ parameterize hadronic
matrix elements.
Therefore, a measurement of the $\bd$ ($\bs$) oscillation frequency allows the
CKM matrix element $V_{td}$ ($V_{ts}$) to be determined. However, Lattice
QCD calculations\cite{Draper} of the product $f_{B_q} \sqrt{B_{B_q}}$ are plagued by an uncertainty of 20-25\%. This uncertainty limits the precision of the extraction of $V_{td}$
from the fairly precise measured value\cite{boscwg}
of the $\bd$ oscillation frequency
$\dmd = 0.476 \pm 0.016$ ps$^{-1}$.
Theoretical uncertainties are significantly reduced in the ratio between
$\bs$ and $\bd$ oscillation frequencies:\cite{Draper}
\begin{equation}
  \frac{\dms}{\dmd} = \frac{m_{B_s} f_{B_s}^2 B_{B_s}}{m_{B_d} f_{B_d}^2 B_{B_d}}
                      \left|\frac{V_{ts}}{V_{td}}\right|^2
                    = (1.14 \pm 0.06)^2 \left|\frac{V_{ts}}{V_{td}}\right|^2~.
\end{equation}
Invoking CKM unitarity to obtain $|V_{ts}|^2$ from the measured value for $|V_{cb}|^2$,
one can then extract $V_{td}$ with good precision.

\begin{figure}[t]
  \vspace*{-14mm}
  \hspace*{6mm}
  \centering
  \epsfxsize=10.5cm
  \epsfbox{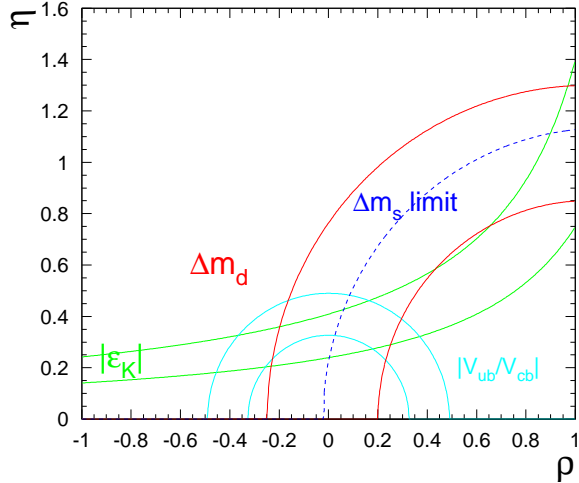}
  \vspace*{-30mm}
  \caption{\it \label{fig_rhoeta}
  \baselineskip=12pt
  Constraints on the apex $(\rho, \eta)$ of the unitarity triangle.
  The area to the left of the dashed line is excluded by the current lower
  limit on $\dms$.
  }
  \baselineskip=18pt
\end{figure}

  Determination of the CKM element $\vtd$ is of great importance since it is
sensitive to the CP violating phase in the Standard Model. In the Wolfenstein
parameterization of the CKM matrix,
$|\vtd|^2 = A^2 \lambda^6 [(1 - \rho)^2 + \eta^2]$
and $|\vts|^2 = A^2 \lambda^4$.
The parameters $\lambda \equiv \sin\theta_c$ and $A$
are well-known but $\rho$ and $\eta$ are not. A non-vanishing value for $\eta$
implies the existence of CP violation in weak decays.
The impact of $\dmd$ and $\dms$ measurements on the knowledge of the
fundamental parameters $\rho$ and $\eta$ is presented in Fig.~\ref{fig_rhoeta},
along with the constraints from the measurement of CP violation in the $\kmix$
system ($\epsilon_K$) and the measurement of $b \to u$ transitions
($|V_{ub} / V_{cb}|$).
From the above parameterization, it is clear that $\bs$ oscillations are
very fast: $\dms / \dmd \simeq 1 / \lambda^2$, which is of order 20.
Resolving these rapid oscillations thus poses a serious experimental challenge.

\section{Past}

The first evidence for $\bmix$ mixing was reported by UA1 in 1987
with a study of like-sign muon pairs
produced in $\overline{p}\, p$ collisions.\cite{UA1}
The rate for like-sign pairs was found to exceed the expected background and
was thus interpreted as evidence (2.9 $\sigma$) for $\bmix$ mixing.
Later in the same year, conclusive evidence was presented by ARGUS
in a study of like-sign dileptons produced
by $e^+ e^-$ annihilation at the $\Upsilon_{4S}$ resonance.\cite{ARGUS}
ARGUS determined the time-integrated
$\bdmix$ mixing probability to be
$\chi_d = 0.17 \pm 0.05$.
Such a large mixing probability came as a surprise and indicated
that the top quark mass had to be very large.

Further studies of like-sign dilepton events produced in $\epem$ annihilation
at the $\Upsilon_{4S}$ and $Z^0$ resonances, as well as in $\overline{p}\, p$
collisions, confirmed the large rate of mixing. In 1993, the world averages
were\cite{Moser93}
$\chi_d = 0.162 \pm 0.021$ ($\Upsilon_{4S}$) and
$\langle \chi \rangle = 0.117 \pm 0.010$ ($Z^0$),
where $\langle \chi \rangle$ represents the average mixing probability
over the different types of $b$ hadrons produced in $Z^0$ decays,
$\langle \chi \rangle = f(\bd)\, \chi_d + f(\bs)\, \chi_s$,
where $f(\bd)$ and $f(\bs)$ are the fractions of $\bd$ and $\bs$ in the
selected sample, respectively. 
Combining these two results, a lower limit on the $\bs$ oscillation frequency was obtained:\cite{Moser93} $\dms > 0.5$ $\ips$ at 90\% C.L.

In 1994, ALEPH extended the like-sign dilepton technique by incorporating
the $B^0$ proper decay time to investigate the time dependence of $\bsmix$
mixing for the first time.
Using a sample of 1 million hadronic $Z^0$ decays,
a direct limit of $\dms > 1.8$ $\ips$ (95\% C.L.) was obtained.\cite{ALEPH94}
Shortly after, OPAL improved the limit to $\dms > 2.2$ $\ips$
in a similar study based on 1.5 million $Z^0$ decays.\cite{OPAL95}
These initial studies were limited mostly by the low efficiency of the dilepton
event selection. Analyses were later improved by incorporating new vertex
selection and tagging algorithms, as described in the next section.

\section{Present}

Studies of time-dependent oscillations require three ingredients:
(i) reconstruction of the $\bs$ decay proper time,
(ii) determination of the $\bs$ or $\bsb$ flavor at production, and
(iii) determination of the flavor at decay.
Decays for which the production and decay flavors are different are tagged
as ``mixed'', otherwise they are tagged as ``unmixed''.
The significance for a $\bs$ oscillation signal can be approximated
by\cite{Moser}
\begin{equation}
  S = \sqrt{\frac{N}{2}}\: f(\bs)\: \left[1 - 2\, w\right]\:
      e^{-\frac{1}{2} (\dms \sigma_t)^2} ,
  \label{eq_signif}
\end{equation}
where $N$ is the total number of decays selected, $w$ is the probability to
incorrectly tag a decay as mixed or unmixed (i.e. the mistag rate)
and $\sigma_t$ is the proper time resolution.
The proper time resolution depends on both the decay length resolution
$\sigma_L$ and the momentum resolution $\sigma_p$ according to
$\sigma_t^2 = (\sigma_L / \gamma\beta c)^2 + (t\, \sigma_p/p)^2$.
The ability to resolve rapid $\bs$ oscillations thus requires excellent
decay length and momentum resolution, and benefits from having
a low mistag rate and a high $\bs$ purity.

Tagging of the production flavor is performed by combining several techniques.
The most powerful technique exploits the large polarized forward-backward
asymmetry of $\Zbb$ decays (available at SLD only).
In this case, a left- (right-) handed
incident electron tags the forward hemisphere quark as a
$b$ ($\overline{b}$) quark.
Other tags used by most experiments rely on charge information from
the hemisphere opposite that of the $\bs$ decay candidate (i.e. the
hemisphere expected to contain the other $b$ hadron in the event):
(i) charge of lepton from the direct transition $b \to l^-$,
(ii) momentum-weighted jet charge,
(iii) secondary vertex charge, and
(iv) charge of kaon from the dominant decay transition $b \to c \to s$.
Information from the same hemisphere is also used:
(i) unweighted (or weighted) jet charge, and
(ii) charge of fragmentation kaon.
The different tags are combined to provide effective mistag rates of
$\sim 25\%$ for LEP
experiments and up to $\sim 15\%$ for SLD.

The various analyses differ mostly in the way $\bs$ decay candidates are reconstructed,
which in turn affects the quality of the decay flavor tag and the $\bs$ purity.
Analyses can be grouped in three main categories: inclusive, semi-exclusive,
and fully exclusive.
Inclusive methods have the advantage of large statistics but suffer from low purity,
whereas more exclusive methods yield small sample sizes but benefit from a much increased sensitivity per event.

\subsection{Inclusive Methods}

Inclusive reconstruction of semileptonic decays has been investigated by ALEPH,
DELPHI, OPAL, and SLD.
The method typically relies on the selection of identified leptons
($e$ or $\mu$) with sufficiently large momentum transverse to the $b$ jet
(the minimum $p_T$ is usually 1 GeV/c) in order to reduce the contribution
from cascade decays ($b \to c \to l^+$). Direct leptons from $b \to l^-$
transitions contribute $\sim 90\%$ of all selected leptons.
As a result, the decay flavor tag is very clean.
The charm decay vertex is reconstructed topologically and the
resultant ``D'' track is
intersected with the lepton trajectory to define the $B$ decay point.

This method benefits from high statistics and a low mistag rate
for the decay flavor tag
but suffers from a low $\bs$ purity (typically 10-15\%).
The sensitivity of the method is enhanced by estimating the mistag rates,
the $\bs$ purity and the proper time resolution event by event.
For example, the most sensitive analysis by ALEPH selects 33023 events, with
an estimated $\bs$ purity of 10.4\% (close to the $\bs$ production fraction).
The sample is divided into 11 subsamples
with $\bs$ purity varying between 5\% and 24\%, depending upon the charm
vertex track multiplicity, the charge and momentum of tracks in the vertex,
as well as the presence of identified kaons in the vertex.

SLD has devised novel inclusive methods relying on the
excellent tracking resolution provided by its CCD pixel vertex detector.
In particular, the lepton+D vertex analysis achieves a decay length resolution
$\sigma_L = 67\,\mu$m (60\% fraction) and $\sigma_L = 273\,\mu$m (40\%).
The Charge Dipole analysis attempts to reconstruct the
charged track topology of $\bs \to D_s^- X$ decays by reconstructing both
secondary (``B'') and tertiary (``D'') vertices.
The charge difference between the B and D vertices $\delta q = Q_D - Q_B$
tags the decay flavor ($\delta q < 0$ for $\bs$ and $\delta q > 0$ for $\bsb$)
with a mistag rate of 21\%.
This rate is considerably larger than that achieved with semileptonic
analyses but it is compensated by the increase in statistics due to the
fully inclusive selection.

\subsection{Semi-Exclusive Methods}

Semi-exclusive methods enhance the sensitivity to $\bs$ oscillations
mostly by improving
the $\bs$ purity and, to a lesser extent, the proper time resolution.
This, however, comes at the cost of much lower efficiency.
ALEPH, CDF, and DELPHI perform partial $\bs$ reconstruction
in the modes $\bs \to D_s^- l^+ \nu_l X$ and $\bs \to D_s^- h^+ X$,
where $h$ represents any charged hadron and the $D_s$ decay is either
fully or partially reconstructed in the modes
$D_s^- \to \phi \pi^-, K^{\ast 0} K^-, K^0 K^-, \phi\pi^-\pi^+\pi^-,
\phi l^- \overline{\nu_l}$, etc.

The most sensitive single analysis performed by DELPHI selects 436 $D_s^- l^+$ events.
Despite the low statistics the analysis is competitive due to
its high $\bs \to D_s^- l^+ \nu_l$ purity, estimated to be $\sim 53\%$,
and its good decay length and momentum resolution,
$\sigma_L = 200\,\mu$m (82\% fraction) and $670\,\mu$m (16\%),
$\sigma_p / p = 0.07$ (82\% fraction) and $0.16$ (16\%).

Analyses reconstructing $D_s^- h^+$ final states benefit from increased
statistics but their overall sensitivity is somewhat reduced due to lower
$\bs$ purity and worse resolution.

\subsection{Exclusive Methods}

DELPHI has performed an exploratory analysis in which
the $\bs$ decays are fully reconstructed in the modes
$\bs \to D_s^- \pi^+, D_s^- a_1^+, {\overline{D^0}} K^- \pi^+$,
and ${\overline{D^0}} K^- a_1^+$, where the $D_s^-$ and $\overline{D^0}$
decays are fully reconstructed.
The analysis selects 44 candidates with an estimated $\bs$ purity of
approximately 50\% and an excellent decay length resolution of
$\sigma_L = 117\,\mu$m (58\% fraction) and $216\,\mu$m (42\%).
The uncertainty in momentum is essentially negligible and thus
the oscillation amplitude is not damped at large proper time.
Despite a high sensitivity per event, the analysis is limited by
the available statistics.
Nevertheless, it is clearly the method of choice for future studies of
$\bsmix$ mixing at hadron colliders (see Sec.~\ref{sec_future}).

\subsection{World Average}

The fit for the $\bs$ oscillation frequency is performed using the
amplitude method.\cite{Moser}
In this method,
the unmixed (mixed) probability density is expressed as
$P(B^0 \to B^0(\overline{B^0})) =
 \frac{\Gamma}{2} e^{-\Gamma t} (1 \pm A\cos\Delta m\, t)$. 
A fit is then performed to determine the oscillation amplitude ``$A$'' at
a series of fixed frequencies. Amplitude values of $A = 0$ are expected
for frequencies sufficiently different from the true oscillation frequency and a value
of $A = 1$ is expected at the true frequency.
The amplitude method is thus similar to a normalized Fourier transform.

\begin{figure}[t]
  \vspace*{-8mm}
  \centering
  \epsfxsize=11cm
  \epsfbox{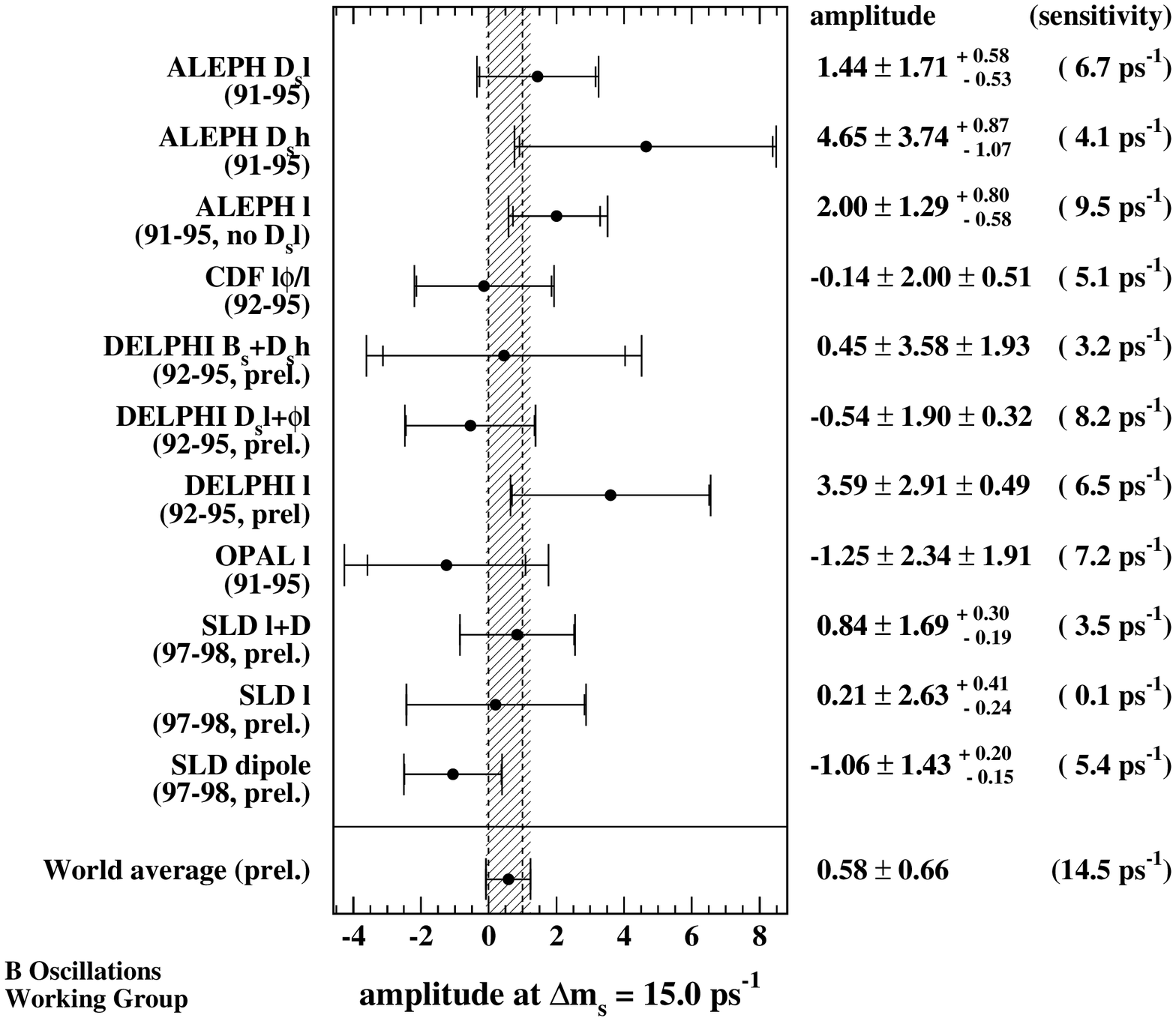}
  \vspace*{-2mm}
  \caption{\it \label{fig_ampcompar}
  \baselineskip=12pt
  Measurements of the $\bs$ oscillation amplitude at $\dms = 15\,\ips$.}
  \baselineskip=18pt
\end{figure}

\begin{figure}[t]
  \vspace*{-7mm}
  \centering
  \epsfxsize=9cm
  \epsfbox{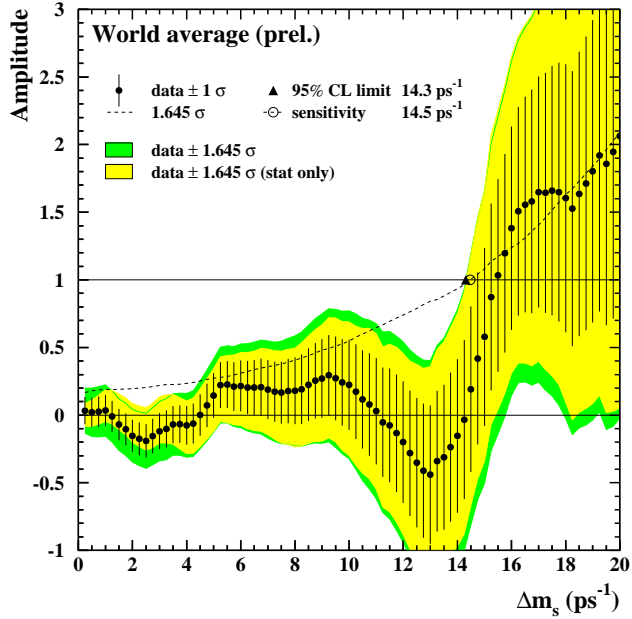}
  \vspace*{-2mm}
  \caption{\it \label{fig_afit_W}
  \baselineskip=12pt
  Measured oscillation amplitude as a function of $\dms$.}
  \baselineskip=18pt
\end{figure}

The measured amplitude at $\dms = 15\,\ips$ for the various analyses
is shown in Fig.~\ref{fig_ampcompar}.
Also shown is the sensitivity of each analysis to set a 95\% C.L. lower limit
on $\dms$.
These analyses have been combined,\cite{boscwg}
taking correlated systematic uncertainties
into account and the resulting world average amplitude spectrum
is shown in Fig.~\ref{fig_afit_W}.
Mixing (A=1) is excluded for $\dms < 14.3\,\ips$ at 95\% C.L.,
a limit close to the expected sensitivity of 14.5 $\ips$
(obtained by setting the measured amplitude to zero).
Beyond that limit, the uncertainties become too large (due to the limited
proper time resolution) to discriminate between mixing and no mixing.

\section{Future}
\label{sec_future}

In the near future, ALEPH, DELPHI, OPAL, and SLD expect to further
improve their sensitivity by adding new analysis techniques and
refining existing analyses.
Beyond LEP and SLD, future experiments at hadron machines are expected to bring
the study of $\bs$ oscillations to a new level.
By exploiting the tremendous cross section for $b$ hadrons at those machines
and designing new trigger schemes aimed at identifying secondary
vertices,
HERA-B expects to observe a $3\,\sigma$ signal
for $\dms$ up to $\sim 27$ $\ips$,
whereas  CDF, BTeV, and LHC-b expect to observe a $5\,\sigma$ signal
for $\dms$ up to about 40 $\ips$, 40 $\ips$, and 48 $\ips$, respectively.
If a signal is found, the statistical precision on $\dms$ is
predicted to be excellent
(e.g., LHC-b expects to achieve a precision better than 0.1\%).

\section*{Acknowledgements}
\vspace{-1mm}
I wish to thank the following people for their help:
G.~Blaylock, R.~Forty, R.~Hawkings,
M.~Paulini, P.~Roudeau, O.~Schneider, and A~.Stocchi.

\end{document}